\documentclass[aps,prl,showpacs,amsmath,amssymb,amsfonts,twocolumn,superscriptaddress,checklength]{revtex4-1}
\usepackage{graphicx}
\usepackage{subfigure}
\usepackage{verbatim}
\usepackage{dcolumn}% Align table columns on decimal point
\usepackage{bm}% bold math
\usepackage{epsf}
\usepackage{xcolor}
\usepackage{hyperref}
\usepackage{hhline}
\usepackage{txfonts}%规范字体
\usepackage{float}
\usepackage{enumerate}
\usepackage{bbm}
\usepackage{lipsum}
\usepackage{amsmath}
\usepackage{amsfonts}
\hypersetup{
	colorlinks=true,   % Options: false (boxed links); true (colored links).
	linkcolor=blue,  % Color of internal links.
	citecolor=blue, % Color of links to bibliography.
	urlcolor =blue    % Color of external links.
}

\begin{document}
\title{Enhanced Strong Coupling between Spin Ensemble and non-Hermitian Topological Edge States}

\author{Jie Qian}
\affiliation{Interdisciplinary Center of Quantum Information, Zhejiang Province Key Laboratory of Quantum Technology and Device, Department of Physics, Zhejiang University, Hangzhou 310027, China}
\author{Jie Li}
\affiliation{Interdisciplinary Center of Quantum Information, Zhejiang Province Key Laboratory of Quantum Technology and Device, Department of Physics, Zhejiang University, Hangzhou 310027, China}
\author{Shi-Yao Zhu}
\affiliation{Interdisciplinary Center of Quantum Information, Zhejiang Province Key Laboratory of Quantum Technology and Device, Department of Physics, Zhejiang University, Hangzhou 310027, China}
\affiliation{Hefei National Laboratory, Hefei 230088, China}
\author{J. Q. You}
\email{jqyou@zju.edu.cn}
\affiliation{Interdisciplinary Center of Quantum Information, Zhejiang Province Key Laboratory of Quantum Technology and Device, Department of Physics, Zhejiang University, Hangzhou 310027, China}

\author{Yi-Pu Wang}
\email{yipuwang@zju.edu.cn}
\affiliation{Interdisciplinary Center of Quantum Information, Zhejiang Province Key Laboratory of Quantum Technology and Device, Department of Physics, Zhejiang University, Hangzhou 310027, China}

\begin{abstract}
Light-matter interaction is crucial to both understanding fundamental phenomena and developing versatile applications. Strong coupling, robustness, and controllability are the three most important aspects in realizing light-matter interactions. Topological and non-Hermitian photonics, have provided frameworks for robustness and extensive control freedom, respectively. How to engineer the properties of the edge state such as photonic density of state, scattering parameters by using non-Hermitian engineering while ensuring topological protection has not been fully studied. Here we construct a parity-time-symmetric dimerized photonic lattice and generate complex-valued edge states via spontaneous PT-symmetry breaking. The enhanced strong coupling between the topological photonic edge mode and magnon mode in a ferromagnetic spin ensemble is demonstrated. Our research reveals the subtle non-Hermitian topological edge states and provides strategies for realizing and engineering topological light-matter interactions. 
\end{abstract}
\maketitle

\textit{Introduction.---}Topology has evolved as a powerful governing principle for predicting and harnessing the robust propagation of currents in various systems, including condensed matter system~\cite{Burkov-16,Hasan-10}, acoustics~\cite{Zhaoju-15,Ma-19,Yihao-22}, mechanics~\cite{Huber-16} and photonics~\cite{Haldane-08,Wang-09,Lu-14,Ozawa-19}. In topological photonics, a topological invariant ensures robust localization or propagation of electromagnetic waves~\cite{Blanco-Redondo-18,Yang-18,Klembt-18}. On the other hand, non-Hermitian photonics~\cite{Feng-17,EI-Ganainy-18,Longhi-18} has also flourished in recent years, not only due to the ubiquitous non-Hermiticity in nature~\cite{Bender-07}, but also because the non-Hermiticity provides additional degrees of freedom to manipulate the wave behaviors. In pursuit of the simultaneous robustness and greater control flexibility, as well as the interest in fundamental research, non-Hermitian topological physics~\cite{Ashida-20,Coulais-21,Bergholtz-21} has received considerable attention and substantial development. Scientists investigate new paradigms~\cite{Yao-18,Yokomizo-19,CHL-20,Helbig-20,Xue-20} and explore potential applications in this interdisciplinary territory~\cite{Zhao-19,St-Jean-17,Parto-18,Hu-21}.
\begin{figure}[t]
	\includegraphics[width=0.47\textwidth]{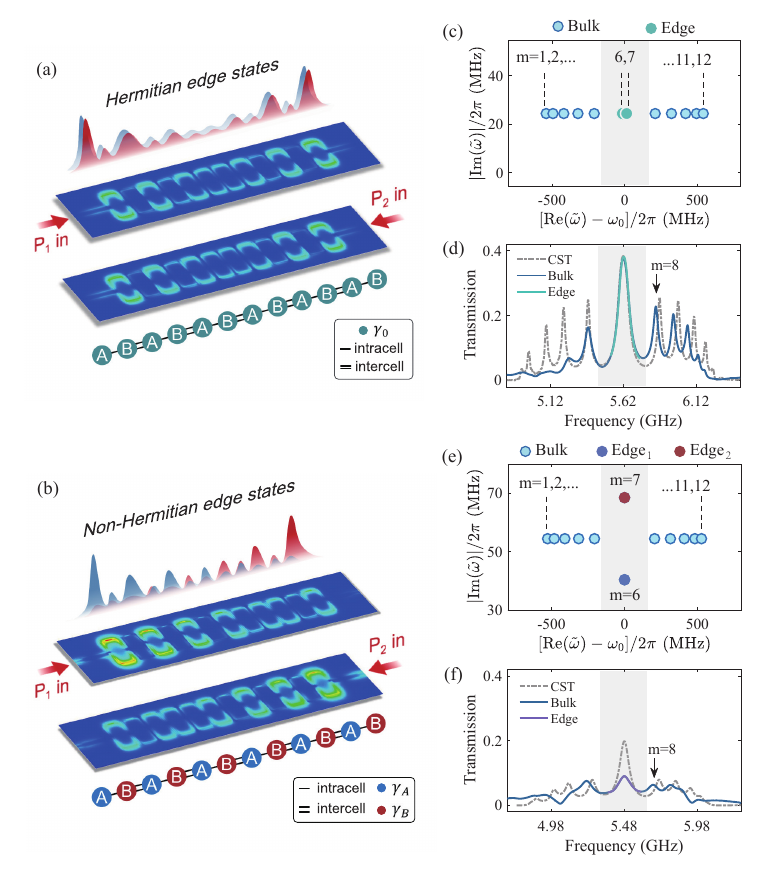}
	\caption{(a)(b) Schematic diagram of the Hermitian and non-Hermitian SSH chains. (c) Eigenmodes of the Hermitian SSH chain are plotted in the complex energy plane. The zero-energy modes exist in the band gap. (d)(f) Transmission spectra of the Hermitian and non-Hermitian SSH chains. (e) Eigenmodes of the non-Hermitian SSH chain are plotted in the complex energy plane. The alternated on-site losses result in spontaneous PT-symmetry breaking of the edge modes.   \label{fig1}}
\end{figure}

\begin{figure*}[t!]
	\centering
	\includegraphics[width=0.8\textwidth]{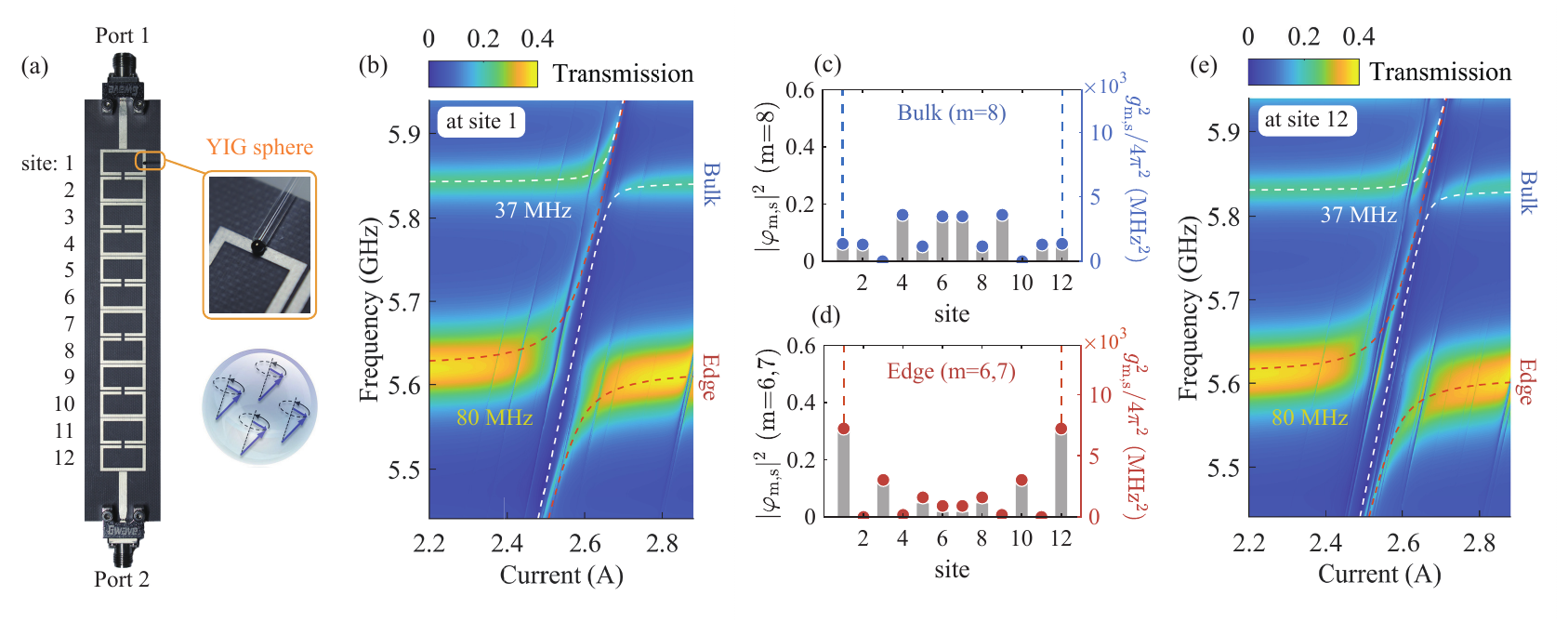}
	\caption{Photograph of the Hermitian SSH chain. The chain contains six unit cells and twelve SRRs. The SRRs are labled by a site index $s$. The YIG sphere is placed on the top of the device. (b)(e) The mappings of the transmission spectra are plotted versus the electromagnet current and probe frequency when a YIG sphere is placed at site-1 and site-12, respectively. Strong coupling between the edge (bulk) mode and the magnon mode is indicated by the large (small) level repulsion. (c) The squares of the coupling strengths $g_{m,s}^2$ ($m$=8) (blue dots) are extracted when the YIG sphere is positioned at the $s$-th site, where $m$ is the eigenmode index. The gray bars represent the intensity distributions of the bulk state wave function $|\varphi_{m,s}|^2$ ($m$=8). (d) The squares of the coupling strengths $g_{m,s}^2$ ($m$=6,7) (red dots) are plotted versus site index $s$. The intensity distributions of the edge state wave functions $|\varphi_{m,s}|^2$ (m=6,7) are depicted by the gray bar.
	}\label{fig2}
\end{figure*}

A coupled system can have two forms of non-Hermiticity. One kind is generated when there is asymmetric interaction between the sites, which leads to the non-Hermitian skin effect~\cite{Yao-18,Alvarez-18}. The other type, which is caused by on-site loss, can lead to intriguing phenomena associated with the parity-time (PT) symmetry. The PT-symmetric systems have received special attention, because they were proved to have real spectra~\cite{Bender-98}. A sequence of studies have studied the topologically protected bound (defect) states in PT-symmetric topological systems~\cite{Schomerus-13,Malzard-15,Weimann-17,Stegmaier-21}, where the defect states are real in the PT-symmetry unbroken phase. Moreover, a number of studies have investigated whether topological edge states exist in the PT-symmetric systems~\cite{Esaki-11,Hu-11,Xue-17,Cheng-22}, concluding that since the edge state is not an eigenstate of the PT operator, an imaginary eigenvalue is obtained along with the spontaneous PT-symmetry breaking. In this case, a non-Hermitian edge state is obtained. We find that these imaginary edge states in the PT-symmetric system are actually topologically protected by the particle-hole symmetry~\cite{SM}.  In the one-dimensional (1D) non-Hermitian PT-symmetric Su-Schrieffer-Heeger (SSH) model~\cite{Su-79}, the chiral symmetry of the system is broken, losing its topological $\mathbb{Z}$ invariant, but the particle-hole symmetry of the system is preserved and the system owns a topological $\mathbb{Z}_2$ invariant. In the presence of perturbations that do not violate the particle-hole symmetry, the real parts of the eigenvalues of the edge modes remain 0, reflecting the topologically protected characteristics. Under this situation, the topological photonic mode with robust properties can be further manipulated by non-Hermiticity, which is highly desirable for investigating light-matter interactions~\cite{Gutzler-21,Ruggenthaler-18,Kockum-19}.

To investigate the interaction between topological photonic modes and matters~\cite{Kim-21}, we employ the photon-magnon coupling system~\cite{Huebl-PRL-2013,Tabuchi-PRL-2013,Zhang-PRL-2014,Tobar-PRApp-2014,You-npj-2015,Wang-2019,Wang-2020,Rameshti-22,Yuan-22}, which has benefits including the flexible tunability and experimental demonstration at room temperature. In this Letter, we use a set of lossy microwave resonators to build 1D non-Hermitian SSH photonic lattices. By coupling a ferromagnetic spin ensemble (FSE) to Hermitian and non-Hermitian SSH chains and monitoring the strength of the coupling between the photonic modes and the magnon mode in the FSE, we verify the topological edge states and bulk states. Non-Hermiticity introduced by the on-site alternating losses breaks the passive PT-symmetry of zero-energy modes and results in two complex-valued edge states, which localize exponentially at the opposite ends of the chain [Fig.~\ref{fig1}(b)]. Further, the photonic density of state (PDOS) at boundaries is larger than that in the Hermitian case [Fig.~\ref{fig1}(a)], which strengthens the coupling between the topological photonic mode and the magnon mode. Our experiment demonstrates the potential of manipulating the interaction between topological photonic states and matter by exploiting non-Hermiticity.

\textit{System and model.---}The SSH chain consists of six unit cells [Figs.~\ref{fig1}(a) and \ref{fig1}(b)], in which each unit contains two split-ring-resonators (SRRs) fabricated on the F4B substrate [Fig.~\ref{fig2}(a)]. In the experiment, the SRR exhibits a resonance at $\omega_0/2\pi$=5.62 GHz with an intrinsic loss of $\gamma_0/2\pi$=24.42 MHz, and the topological property is unaltered by the uniform losses along the chain~\cite{Feng-17}. Therefore, SRRs with the same loss can be used to build the Hermitian SSH model. Two neighboring SRRs are separated by staggered spacings to realize the intracell and intercell coupling rates, $v$ and $w$. Edge states appear in the finite chain when the bulk winding number of the Hermitian Hamiltonian is $\mathcal{W}_{\rm{h}}$=1~\cite{Weimann-17}. The effective Hermitian SSH chain is designed in the topological non-trivial phase ($v/2\pi$=216.5 MHz, $w/2\pi$=341 MHz) and the Hamiltonian is written as~\cite{SM}:
\begin{equation} 
\mathcal{H}_{\rm{h}}/{\hbar}=\sum_{s=1}^{2N}(\omega_{0}-i\gamma_{0})\hat{a}_{s}^{\dagger}\hat{a}_{s}+\sum_{s=1}^{2N-2}(v\hat{a}_{s}\hat{a}_{s+1}^{\dagger}+w\hat{a}_{s+1}\hat{a}_{s+2}^{\dagger}),
\end{equation}
where ${\hat{a}}_s^\dag~({\hat{a}}_s)$ is the photon creation (annihilation) operator of the $s$-th SRR. The uniform losses of the units only yield all eigenvalues of the chain to have the same imaginary component $i\gamma_0$. The eigenvalues of the coupled SRRs are plotted in the complex plane, as shown in Fig.~\ref{fig1}(c). A pair of zero-energy modes (Re(${\widetilde{\omega}_{m=6,7}})-\omega_{0}$=0, green dots) appear in the band gap (gray area), which are the edge modes. The measured transmission spectrum of the chain is shown in Fig.~\ref{fig1}(d), where the peaks correspond to the resonances of the eigenmodes. By simulating the field distribution at the edge mode frequency of $\omega_0/2\pi$=5.62 GHz, we find that the electromagnetic field tends to localize at both edges of the chain, as predicted by wave function distribution~\cite{SM}. In the low-frequency region, the measured spectrum [Fig.~\ref{fig1}(d), solid line] displays an amplitude deviation from that in the high-frequency region. This is due to the residual dissipative coupling between SRRs~\cite{Weimann-17,SM}. 
 
\begin{figure*}[t]
	\includegraphics[width=0.8\textwidth]{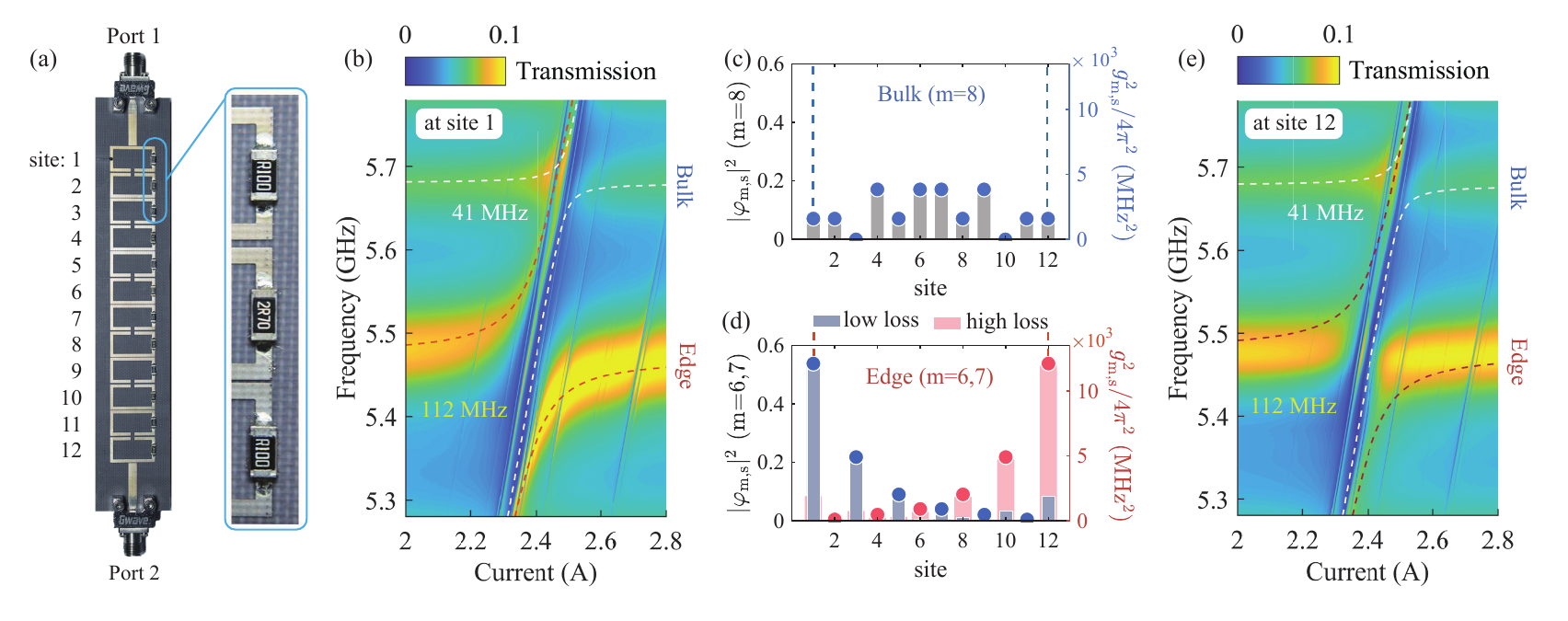}
	\caption{Photograph of the non-Hermitian SSH chain. The integrated resistors induce on-site losses. (b)(e) When the YIG sphere is placed at site-1 and site-12, respectively, the mappings of the transmission spectra are plotted versus the electromagnet current and probe frequency. The resonances with frequencies of 5.48 GHz and 5.68 GHz correspond to the non-Hermitian edge mode ($m$=6,7) and bulk mode ($m$=8), respectively. (c) The squares of the coupling strengths $g_{m,s}^2$ ($m$=8) (blue dots) are plotted versus site index $s$. Gray bars represent the intensity distributions of the bulk state wave function $|\varphi_{m,s}|^2$ ($m$=8). (d) The squares of the coupling strengths $g_{6,s}^2$ (blue dots) and $g_{7,s}^2$ (red dots) are plotted versus site index $s$. The intensity distributions of edge state wave functions $|\varphi_{m,s}|^2$ are denoted by gray ($m$=6, low loss edge state) and red ($m$=7, high loss edge state) bar diagrams, respectively. }\label{fig3}
\end{figure*}

Then, on-site non-Hermiticity is added to the SSH chain. As depicted in Fig.~\ref{fig3}(a), resistors $R_{\rm{A}}=0.1~\Omega$ and $R_{\rm{B}}=2.7~\Omega$ are integrated into odd and even sites of the chain, respectively, which induce alternated losses of $\gamma_{\rm{A}}/2\pi$=36 MHz and $\gamma_{\rm{B}}/2\pi$=73 MHz. The Hamiltonian becomes~\cite{SM}:
\begin{equation}\label{eq2}
\begin{split}
		\mathcal{H}_{\rm{nh}}/{\hbar}=&\sum_{s\in X}(\omega_{0}-i\gamma_{\rm{A}})\hat{a}_{s}^{\dagger}\hat{a}_{s}+\sum_{s\in Y}(\omega_{0}-i\gamma_{\rm{B}})\hat{a}_{s}^{\dagger}\hat{a}_{s}
		\\&+\sum_{s=1}^{2N-2}(v\hat{a}_{s}\hat{a}_{s+1}^{\dagger}+w\hat{a}_{s+1}\hat{a}_{s+2}^{\dagger}),
\end{split} 
\end{equation}
where $X$=\{1, 3, 5, ..., 2$N$-1\}, $Y$=\{2, 4, 6, ..., 2$N$\}, and $N$=6. The integrated resistors shift $\omega_0/2\pi$ to 5.48 GHz, and the hopping rates shift to $v/2\pi$=208.5 MHz, and $w/2\pi$=335.5 MHz. The alternated losses make the system a passive PT-symmetric one. The spontaneous PT-symmetry breaking occurs in zero-energy modes, resulting in a splitting of the imaginary parts of zero-energy modes, as shown in Fig.~\ref{fig1}(e). One with a low loss Im(${\widetilde{\omega}}_{m=6})/2\pi$=40.42 MHz ($\rm{Edge}_1$, blue dot) localizes at the left boundary of the chain, and the other with a high loss Im(${\widetilde{\omega}}_{m=7})/2\pi$=68.58 MHz ($\rm{Edge}_2$, red dot) localizes at the right, as schematically shown in Fig.~\ref{fig1}(b). The bulk Hamiltonian still preserves the PT-symmetry when $\delta\gamma/2<\left|w-v\right|$, and $\delta\gamma=\gamma_{\rm{B}}-\gamma_{\rm{A}}$. In this regime, the topological property is still determined by the generalized integer winding number $\mathcal{W}_{\rm{nh}}$~\cite{SM}. $\mathcal{W}_{\rm{nh}}=1$ guarantees the existence of two non-Hermitian topological edge modes.  

\textit{Experiment results.---}To investigate the edge modes engineered by the non-Hermiticity, we measure the PDOS and linewidths of the edge and bulk modes in both Hermitian and non-Hermitian cases. Notably, conventional detection of the PDOS relies on the near-field radiation~\cite{Bellec-13}, but in the non-Hermitian situation, the local gain and loss will diminish its reliability. Using the spin ensemble as a probe, we can directly detect the PDOS. In addition, it allows us to study the strong coherent interaction between the topological photonic modes and magnons.

In the experiment, the spin ensemble employed to couple with the chain is a 1-mm diameter yttrium iron garnet (YIG) sphere. The magnon mode in the sphere interacts with the local photonic modes, with a coupling strength $g$ proportional to $\eta\chi\sqrt{nS\hbar\omega_r/2V}$~\cite{Tabuchi-PRL-2013,Zhang-PRL-2014}, where $\eta\le1$ describes the spatial overlap and polarization matching between the photonic mode and the magnon mode, $\chi$ is the gyromagnetic ratio, $n$ is the total number of spins, $S$=5/2 is the spin number of the ground state ${\rm Fe}^{3+}$ ion in YIG, $\omega_r$ is the resonance frequency, and $V$ is the photonic mode volume. Consequently, the square of the coupling strength $g^2$ directly reflects the PDOS at the coupling location. Firstly, we move the YIG sphere to each site (labeled as $s$, $s$=1,2,3,...,12) of the Hermitian chain, and obtain the PDOS distribution of the $m$-th eigenmode by analyzing the transmission spectra. The bias magnetic field is perpendicular to the device plane, and mappings of transmission spectra are measured versus electromagnet current and probe frequency. Figures~\ref{fig2}(b) and \ref{fig2}(e), for instance, show the mappings when the YIG sphere is placed at site-1 and site-12, respectively. The coupling strength between $m$-th eigenmode of the chain and the magnon mode at the $s$-th site is defined as $g_{m,s}$, which can be obtained by fitting the level repulsion with:
\begin{equation}\label{eq3}
{\widetilde{\omega}}_{m,s}^\pm=\frac{1}{2}\left[{\widetilde{\omega}}_n+{\widetilde{\omega}}_m\pm\sqrt{{{(\widetilde{\omega}}_n-\widetilde{\omega}}_m)+4g_{m,s}^2}\right],	
\end{equation}
where ${\widetilde{\omega}}_n=\omega_n-i\gamma_n$ and ${\widetilde{\omega}}_m=\omega_m-i(\gamma_m+\kappa_m)$ are the eigenvalues of the uncoupled magnon mode and the $m$-th eigenmode of the chain, respectively. $\gamma_n$ is the total loss rate of the magnon mode, $\gamma_m$ is the intrinsic loss rate of the $m$-th eigenmode, and $\kappa_m$ is the extrinsic loss rate of the $m$-th eigenmode to the input/output ports~\cite{SM}. Coupling strengths between the magnon mode and edge modes ($m$=6,7) at site-1 and site-12 are obtained by fitting the level repulsion depicted in Figs.~\ref{fig2}(b) and \ref{fig2}(e), which are $g_{\rm{edge},1}/2\pi=g_{\rm{edge},12}/2\pi=80$ MHz. Similarly, coupling strengths between the magnon mode and bulk mode ($m$=8) at site-1 and site-12 are obtained as $g_{\rm{bulk},1}/2\pi=g_{\rm{bulk},12}/2\pi=37$ MHz. $g_{m,s}^2$ as a function of the site index $s$ are illustrated in Figs.~\ref{fig2}(c) and  \ref{fig2}(d), denoted by blue ($m$=8) and red dots ($m$=6,7), respectively. The observed $g_{m,s}^2$ are in good agreement with the intensity distributions for the wave function $|\varphi_{m,s}|^2$ (gray bar diagram).

\begin{figure}[t!]
	\centering
	\includegraphics[width=0.47\textwidth]{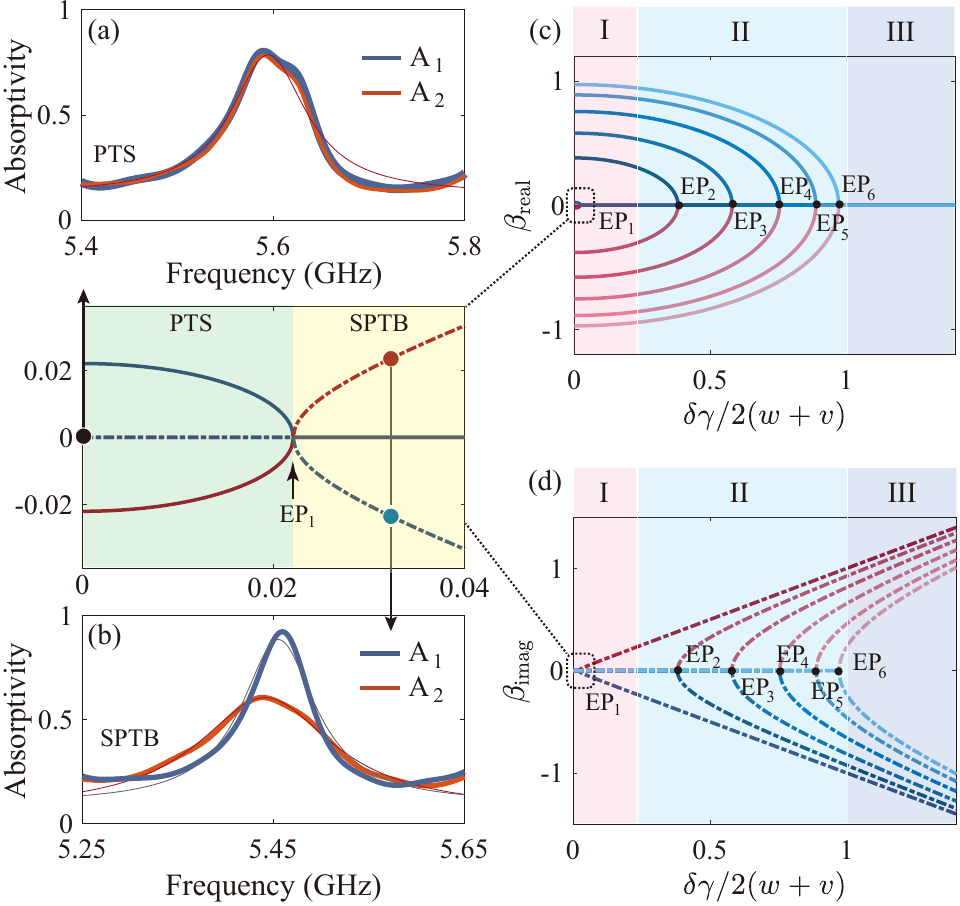}
	\caption{(a)(b) The measured absorptivity spectra for both Hermitian and non-Hermitian SSH chains, where $\delta\gamma/2(w+v)=0$ and 0.034, respectively. A$_1$ (A$_2$) is measured when loading signal to port 1 (2), as shown by the blue (red) line. (c)(d) The real and imaginary parts of normalized eigenvalues, which are plotted versus the non-Hermiticity $\delta\gamma/2(w+v)$. The left inset figure between (a) and (b) shows the PT-symmetry spontaneous breaking point (EP$_1$) of the edge states.}\label{fig4}
\end{figure}

Then, we couple the spin ensemble to the non-Hermitian SSH chain, as shown in Fig.~\ref{fig3}(a). Figures~\ref{fig3}(b) and \ref{fig3}(e) display the mappingswhen the YIG sphere is placed at site-1 and site-12, respectively. The mappings show similar amount of level repulsion, but reflects very different linewidths of the edge modes. Using Eq.~(\ref{eq3}), the loss of the edge mode at site-1 is fitted to be $\gamma_{\rm{edge},1}/2\pi=41.1$ MHz, which is contributed by the addition of the two edge modes ($m$=6,7). The relation is $\gamma_{{\rm{edge}},s}=[{\rm{Im}}({\widetilde{\omega}}_{m=6})\cdot|\varphi_{6,s}|^2+{\rm{Im}}({\widetilde{\omega}}_{m=7})\cdot|\varphi_{7,s}|^2]/(|\varphi_{6,s}|^2+|\varphi_{7,s}|^2)$, and the wave functions of the edge modes $|\varphi_{m,s}|^2$ are displayed as the bar diagram in Fig.~\ref{fig3}(d). Similarly, we get $\gamma_{edge,12}/2\pi$=67.9 MHz. More interestingly, the coupling strengths between the magnon mode and edge modes at site-1 and site-12 are observed to be $g_{\rm{edge},1}/2\pi=g_{\rm{edge},12}/2\pi$=112 MHz, which is larger than that in the Hermitian case (80 MHz). We plot $g_{m,s}^2$ versus site index $s$ for $m$=8 and $m$=6, 7 in Figs.~\ref{fig3}(c) and \ref{fig3}(d), respectively. It can be found that the bulk mode maintains expanded, similar to the Hermitian bulk mode. But, as shown in Fig.~\ref{fig3}(d), the low-loss edge state (Edge$_1$) accumulates at the left boundary, while high-loss edge state (Edge$_2$) accumulates at the right edge. The introduction of on-site loss does contribute to the increase of PDOS at the boundaries. The mechanism can be interpreted as follows: When the PT-symmetry of the edge states is broken, the energy flow between adjacent resonators is partly blocked~\cite{Peng-14}. The low-loss (high-loss) edge state becomes more localized at the low-loss (high-loss) site, as shown in Figs.~\ref{fig1}(b) and \ref{fig3}(a), it corresponds the left (right) boundary of the chain.

It is also intriguing to detect the properties of the non-Hermitian topological edge states from spectroscopic measurements. In the PT-symmetry unbroken phase, two topological edge states cannot be distinguished via spectroscopic measurement, as shown in Fig.~\ref{fig4}(a). The absorptivity spectra $A_1$ measured when loading microwave to port 1 is totally coincident with $A_2$ measured when loading microwave to port 2. In the symmetry broken phase, two topological edge states can be distinguished in spectra, as shown in Fig.~\ref{fig4}(b). The spectra $A_1$ exhibits the low-loss state with a relatively narrow bandwidth, while the spectra $A_2$ reveals the high-loss state. 

Finally, we anticipate to discuss about some additional characteristics of the exceptional point (EP) in the non-Hermitian chain. The dimensionless eigenvalues are defined as $ \beta_{\rm{real}}+i\beta_{\rm{imag}}$, where $\beta_{\rm{real}}=\left[\rm{Re}(\widetilde{\omega})-\omega_0\right]/(v+w)$, $\beta_{\rm{imag}}=\left[|\rm{Im}(\widetilde{\omega})|-\bar{\gamma}\right]/(v+w)$, and $\bar{\gamma}=(\gamma_{\rm{A}}+\gamma_{\rm{B}})/2$. In a finite SSH chain, when increasing the non-Hermitian parameter $\delta\gamma/2(v+w)$, a series of exceptional points are gradually reached [Figs.~\ref{fig4}(c) and \ref{fig4}(d)]. It can be found that the EP of the edge modes is distinctly away from the EPs of the bulk modes. The edge modes experience spontaneous PT-symmetry breaking (SPTB) at EP$_1$, where $\delta\gamma/2(v+w)$ is only about 0.02. With the increase of chain length, the non-Hermiticity needed for SPTB in edge modes decreases exponentially. In the case of $N\gg1$, any finite $\delta\gamma$ will lead to the SPTB in edge modes~\cite{SM}. However, the minimum requirement of SPTB in bulk mode needs $\delta\gamma/2\textgreater\left|w-v\right|$, which is much larger than 0.02. Additional analysis is provided in the supplementary materials.

\textit{Conclusion.---}We have implemented the PT-symmetric non-Hermitian topological SSH model with microwave resonators and achieved the control of topological edge states using the on-site non-Hermiticity. Through spontaneous PT-symmetry breaking, we obtain the non-Hermitian edge modes, where the photonic mode densities are enhanced at both ends of the chain. We realize the strong coupling between the edge modes and the magnon mode in both Hermitian and non-Hermitian cases. We experimentally verify that the coupling strength between the non-Hermitian edge states and the spin ensemble is stronger than that in the Hermitian situation. Our research illustrates non-Hermiticity engineered topological edge states and paves a way for studying strong coherent interaction between topological photonic modes and matter.

\begin{acknowledgments}
This work is supported by the National Key Research and Development Program of China (No.~2022YFA1405200), National Natural Science Foundation of China (No.~$92265202$, No.~$11934010$, No.~${\rm U}1801661$, and No.~$12174329$), and the Fundamental Research Funds for the Central Universities (No.~$2021{\rm FZZX}001$-$02$).
\end{acknowledgments}

\end{document}